\documentclass[aps,prd,twocolumn,nofootinbib,showpacs]{revtex4}

\usepackage{graphicx,color}
\usepackage[colorlinks=true, linkcolor=blue, citecolor=blue, urlcolor=blue]{hyperref}

\begin{document}

\title{ \begin{center}  PMC$_\infty$: Infinite-Order Scale-Setting using the Principle of Maximum Conformality
\\ A Remarkably Efficient Method for Eliminating Renormalization Scale Ambiguities\\ for Perturbative QCD \
\end{center} }

\author{Leonardo Di Giustino$^{1,2}$}
\email[email:]{leonardo.digiustino@gmail.com}
\author{Stanley J. Brodsky$^2$ }
\email[email:]{sjbth@slac.stanford.edu}
\author{ Sheng-Quan Wang$^{2,3}$}
\email[email:]{sqwang@cqu.edu.cn}
\author{ Xing-Gang Wu$^4$}
\email[email:]{wuxg@cqu.edu.cn}

\address{$^1$Department of Science and High Technology, University of Insubria, via valleggio 11, I-22100, Como, Italy}

\address{$^2$SLAC National Accelerator Laboratory, Stanford University, Stanford, California 94039, USA}

\address{$^3$Department of Physics, Guizhou Minzu University, Guiyang 550025, P.R. China}
\address{$^4$Department of Physics, Chongqing University, Chongqing 401331, P.R. China}

\date{\today}

\begin{abstract}

We identify a property of renormalizable SU(N)/U(1) gauge
theories, the {\it Intrinsic Conformality} ($iCF$), which
underlies the scale invariance of physical observables and leads
to a remarkably efficient method to solve the conventional
renormalization scale ambiguity at every order in pQCD:
the~PMC$_\infty$. This new method reflects the underlying
conformal properties displayed by pQCD at NNLO, eliminates the
scheme dependence of pQCD predictions and is consistent with the
general properties of the PMC (Principle of Maximum Conformality).
We introduce a new method to identify conformal and $\beta$-terms
which can be applied either to numerical or to theoretical
calculations. We illustrate the PMC$_\infty$ for the thrust and
C-parameter distributions in $e^+ e^-$ annihilation and then we
show how to apply this new method to general observables in QCD.
We point out how the implementation of the PMC$_\infty$ can
significantly improve the precision of  pQCD predictions; its
implementation in multi-loop analysis also simplifies the
calculation of higher orders corrections in a general
renormalizable gauge theory.

\pacs{11.15.Bt, 11.10.Gh,12.38.Bx,13.66.Jn,13.87.-a}

\end{abstract}

\maketitle

\subsection{Introduction}

A key issue in making precise predictions in QCD is the
uncertainty in setting the renormalization scale $\mu_R$ in order
to determine the correct running coupling $\alpha_s(\mu_R^2)$ in
the perturbative expansion of a scale invariant quantity.
The "Conventional" practice of simply guessing the scale $\mu_R$
of the order of a typical momentum transfer Q in the process, and
then varying the scale over a range Q/2 and 2Q,  gives predictions
which depend on the renormalization scheme, retains dependence on
the initial scale choice, leads to a non-conformal series which
diverges as $\sim \alpha_s^{n} \beta_0^n n!$,  and is even invalid
for QED\cite{maitre:2009xp}. In fact, a physical process will
depend on many invariants and thus have multiple renormalization
scales which depend on the dynamics of the process. Other
proposals for renormalization scale setting such as
PMS~\cite{stevenson} or FAC~\cite{grunberg} not only have the same
difficulties, but they also lead to incorrect and unphysical
results~\cite{kramer}.
 It has been shown recently how all the theoretical constraints
 can be satisfied at once, leading to accurate results by using the PMC~(the Principle of
Maximum Conformality)~\cite{PMC1,PMC2,PMC3}. The primary purpose
of the PMC method is to solve the scale-setting ambiguity, it has
been extended to all orders~\cite{mojaza1,mojaza2} and it
determines the correct running coupling and the correct momentum
flow accordingly to the RGE invariance~\cite{xing1,xing2}. This
leads to results that are invariant respect to the initial
renormalization scale in agreement with the requirement of scale
invariance of an observable in pQCD~\cite{xing3}.\newline We show
here, how the implementation at all orders of the PMC simplifies
in many cases by identifying only the $\beta_0$-terms at each
order of accuracy due to the presence of a new property of the
perturbative corrections. First we recall that there is no
ambiguity in setting the renormalization scale in QED. The
standard Gell-Mann-Low scheme determines the correct
renormalization scale identifying the scale with the virtuality of
the exchanged photon~\cite{qed}. For example, in electron-muon
elastic scattering, the renormalization scale is the virtuality of
the exchanged photon, i.e. the spacelike momentum transfer squared
$\mu_R^2 = q^2 = t$. Thus
\begin{equation} \alpha(t) = {\alpha(t_0) \over 1 - \Pi(t,t_0)}
\label{qed1}
\end{equation} where
$$ \Pi(t,t_0) = {\Pi(t) -\Pi(t_0)\over 1-\Pi(t_0) } $$
From Eq.\ref{qed1} it follows that the renormalization scale
$\mu_R=t$ can be determined by the $\beta_0$-term at the lowest
order. This scale is sufficient to sum all the vacuum polarization
contributions into the dressed photon propagator, both proper and
improper at all orders. Again in QED, considering the case of two
photon exchange a new different scale is introduced in order to
absorb all the $\beta$-terms related to the new subset/subprocess
into the scale. Also in this case the scale can be determined by
identifying the lowest order $\beta_0$-term alone. This term
identifies the virtuality of the exchanged momenta causing the
running of the scale in that subprocess. This scale again would
sum all the contributions related to the $\beta$-function into the
renormalization scale and no further corrections need to be
introduced to the scale at higher orders. Given that the pQCD and
pQED predictions match analytically in the $N_C\to 0 $ limit where
$C_F\alpha_{QCD} \to \alpha_{QED}$ (see ref.~\cite{huet}) we
extend the same procedure to pQCD. In fact in many cases in QCD
 the $\beta_0$ terms alone can determine the pQCD
renormalization scale at all orders~\cite{padeapp} eliminating the
renormalon contributions $\alpha_s^{n} \beta_0^n n!$. Though in
non-Abelian theories other diagrams related to the three- and
four-gluon vertices arise, these terms do not necessarily spoil
this procedure. In fact, in QCD, the $\beta_0$ terms arising from
the renormalization of the three-gluon or four-gluon vertices as
well as from gluon wavefunction renormalization determine the
renormalization scales of the respective diagrams and no further
corrections to the scales need to be introduced at higher orders.
In conclusion, if we focus on a particular class of diagrams we
can fix the PMC scale by determining the $\beta_0$-term alone and
we show this to be connected to the general scale invariance of an
observable in a gauge theory.
 We introduce in this article a parametrization of the observables which stems directly from the
analysis of the perturbative QCD corrections and which reveals
interesting properties like scale invariance independently from
the process or from the kinematics. We point out that this
parametrization can be an intrinsic general property of gauge
theories and we define this property as \textit{Intrinsic
Conformality} (\textit{iCF}{\footnote{Here the Conformality has to
be intended as RG invariance only.}}). We also show how this
property directly indicates what is the correct renormalization
scale $\mu_R$ at each order of calculation and we define this new
method
 PMC$_\infty$:{\it Infinite-Order Scale-Setting using the Principle of Maximum Conformality}.
We discuss the iCF property and the PMC$_\infty$ for the case of
the thrust and C-parameter distributions in $e^+ e^-\rightarrow
3jets$ and we show the results.

In general a normalized IR safe single variable observable, such
as the thrust distribution for the $e^+ e^-\rightarrow  3jets$
\cite{thrust1,thrust2}, is the sum of pQCD contributions
calculated up to NNLO at the initial renormalization scale
$\mu_0$:

\begin{eqnarray}
\frac{1}{\sigma_{0}} \! \frac{O d \sigma(\mu_{0})}{d O}\! &\!\! =
\!\! & \! \left\{ \! \frac{\alpha_{s}(\mu_0)}{2 \pi} \! \frac{ O d
A_{\mathit{O}}(\mu_0)}{d O}\! +\!\!
\left(\!\frac{\alpha_{s}(\mu_{0})}{2 \pi}\!\right)^{\!\!2} \!\!
\frac{ O d B_{\mathit{O}}(\mu_0)}{d O} \right. \nonumber  \\
 & & + \left. \left(\frac{\alpha_{s}(\mu_0)}{2
\pi}\right)^{\!\!3} \! \frac{O dC_{\mathit{O}}(\mu_0)}{d O}\!+
\!{\cal O}(\alpha_{s}^4) \right\},
 \label{observable1}
\end{eqnarray}

where the $\sigma_0$ is tree-level hadronic cross section, the
$A_O, B_O, C_O$ are respectively the LO, NLO and NNLO
coefficients, $O$ is the selected non-integrated variable.
 For sake of simplicity we will refer to the differential coefficients as to \textit{implicit
coefficients} and we drop the derivative symbol, i.e.
\begin{eqnarray}
A_O(\mu_0)~\!\!& \equiv &~~\frac{ O d A_{O}(\mu_0)}{d O},\,
B_O(\mu_0)~ \equiv~~\frac{ O d B_{O}(\mu_0)}{d O}, \nonumber \\
C_O(\mu_0)~\!\! & \equiv &~\frac{ O d C_{O}(\mu_0)}{d O}.
\label{implicitcoefficients}\end{eqnarray}

\medskip We define here the \textit{Intrinsic Conformality} as the
property of a renormalizable SU(N)/U(1) gauge theory, like QCD,
which yields to a particular structure of the perturbative
corrections that can be made explicit representing the
perturbative coefficients using the following
parametrization{\footnote{We are neglecting here other running
parameters such as the mass terms.}}:
\begin{eqnarray}
 A_{O}(\mu_0)\!\!\! &=& \!\!\! A_{\mathit{Conf}} , \nonumber \\
B_{O}(\mu_0) \!\!\! &=& \!\!\! B_{\mathit{Conf}}+\frac{1}{2} \beta_{0} \ln \left(\frac{\mu_0^{2}}{\mu_{I}^{2}}\right) A_{\mathit{Conf}},  \nonumber \\
 C_{O}(\mu_0)\!\!\! &=& \!\!\! C_{\mathit{Conf}} +\beta_{0} \ln \left(\frac{\mu_{0}^{2}}{\mu_{II}^{2}}\right)B_{\mathit{Conf}}+  \nonumber \\
 \!\!\! & & \!\!\! +\frac{1}{4}\left[\beta_{1}+\beta_{0}^{2}
 \ln \left(\frac{\mu_0^{2}}{\mu_{I}^{2}}\right)\right] \ln \left(\frac{\mu_0^{2}}{\mu_{I}^{2}}\right)
 A_{\mathit{Conf}}\nonumber \\
 \label{newevolution}
  \end{eqnarray}
where the $A_{\mathit{Conf}}, B_{\mathit{Conf}},
C_{\mathit{Conf}}$ are the scale invariant \textit{Conformal
Coefficients} (i.e. the coefficients of each perturbative order
not depending on the scale $\mu_0$) while we define the $\mu_N$ as
\textit{Intrinsic Conformal Scales} and $\beta_0,\beta_1$ are the
first two coefficients of the $\beta$-function.  We remind that
the implicit coefficients are defined at the scale $\mu_0$ and
that they change according to the standard RG equations under a
change of the renormalization scale according to :
\begin{eqnarray}
 A_{O}(\mu_R)\!\!\! &=& \!\!\! A_{O}(\mu_0) , \nonumber \\
B_{O}(\mu_R) \!\!\! &=& \!\!\! B_{O}(\mu_0)+\frac{1}{2} \beta_{0} \ln \left(\frac{\mu_R^{2}}{\mu_{0}^{2}}\right) A_{O}(\mu_0),  \nonumber \\
 C_{O}(\mu_R)\!\!\! &=& \!\!\! C_{O}(\mu_0) +\beta_{0} \ln \left(\frac{\mu_{R}^{2}}{\mu_{0}^{2}}\right)B_{O}(\mu_0)+  \nonumber \\
 \!\!\! & & \!\!\! +\frac{1}{4}\left[\beta_{1}+\beta_{0}^{2}
 \ln \left(\frac{\mu_R^{2}}{\mu_{0}^{2}}\right)\right] \ln \left(\frac{\mu_R^{2}}{\mu_{0}^{2}}\right)
 A_{O}(\mu_0)\nonumber \\
 \label{standardevolution}
  \end{eqnarray}
It can be shown that the form of Eq.\ref{newevolution} is scale
invariant and it is preserved under a change of the
renormalization scale from $\mu_0$ to $\mu_R$ by standard RG
equations Eq.\ref{standardevolution}, i.e.:
\begin{eqnarray}
 A_{O}(\mu_R)\!\!\! &=& \!\!\! A_{\mathit{Conf}} , \nonumber \\
B_{O}(\mu_R) \!\!\! &=& \!\!\! B_{\mathit{Conf}} +\frac{1}{2} \beta_{0} \ln \left(\frac{\mu_R^{2}}{\mu_{I}^{2}}\right) A_{\mathit{Conf}},  \nonumber \\
 C_{O}(\mu_R)\!\!\! &=& \!\!\! C_{\mathit{Conf}} +\beta_{0} \ln \left(\frac{\mu_{R}^{2}}{\mu_{II}^{2}}\right)B_{\mathit{Conf}}+  \nonumber \\
 \!\!\! & & \!\!\! +\frac{1}{4}\left[\beta_{1}+\beta_{0}^{2}
 \ln \left(\frac{\mu_R^{2}}{\mu_{I}^{2}}\right)\right] \ln \left(\frac{\mu_R^{2}}{\mu_{I}^{2}}\right) A_{\mathit{Conf}} \label{evolved}
  \end{eqnarray}
We notice that the form of Eq. \ref{newevolution} is invariant and
that the initial scale dependence is exactly removed by $\mu_R$.
Extending this parametrization to all orders we achieve a scale
invariant quantity: \textit{the iCF-parametrization is a
sufficient condition in order to obtain a scale invariant
observable}.

In order to show this property we collect together the terms
identified by the same \textit{conformal coefficient}, we name
each set as \textit{conformal subset} and we extend the property
to the order $n$:
\begin{eqnarray}
& \sigma_I &\!\! =\!\!\left\{ \left(\frac{\alpha_{s}(\mu_{0})}{2
\pi}\right) + \frac{1}{2} \beta_{0} \ln \left(
\frac{\mu_0^{2}}{\mu_{I}^{2}}\right)\left(\frac{\alpha_{s}(\mu_0)}{2
\pi}\right)^2  \right.  \nonumber \\ \!\!\! &\!\!\!\!\! +
\!\!\!\!\!& \!\!\!\left. \!\!\!\! \frac{1}{4}\! \left[ \beta_{1}
\! + \! \beta_{0}^{2} \ln
\left(\frac{\mu_0^{2}}{\mu_{I}^{2}}\right) \right] \ln
\left(\frac{\mu_0^{2}}{\mu_{I}^{2}}\right)\!\!
\left(\frac{\alpha_{s}(\mu_{0})}{2 \pi}\right)^3 \!\!\!+\ldots\!\!
\right\}
\!\! A_{\mathit{Conf}}   \nonumber \\
&  \sigma_{II} &\!\! = \!\! \left\{
\left(\frac{\alpha_{s}(\mu_{0})}{2 \pi}\right)^2 \!\!\!\! +\!
\beta_0 \ln \! \left( \! \frac{\mu_0^2}{\mu_{II}^2}\right)\!\!
\left(\frac{\alpha_{s}(\mu_{0})}{2 \pi}\right)^3 \!\!\! +\ldots
\!\! \right\}\!
B_{\mathit{Conf}}   \nonumber \\
& \sigma_{III} &\!\!= \!\! \left\{
\left(\frac{\alpha_{s}(\mu_{0})}{2
\pi}\right)^3 +\ldots \right\}C_{\mathit{Conf}},   \nonumber \\
& \vdots & \qquad
\qquad  .^{.^{.}} \nonumber \\
&  \sigma_{n} &\!\!= \!\!\left\{
\left(\frac{\alpha_{s}(\mu_{0})}{2 \pi}\right)^n
\right\}\mathcal{L}_{n \mathit{Conf}}, \label{confsubsets}
\end{eqnarray}
 in each subset we have only one intrinsic scale and
only one conformal coefficient and the subsets are disjoint, then
no mixing terms among the scales or the coefficients are
introduced in this parametrization. Besides the structure of the
subsets remains invariant under a global change of the
renormalization scale as shown from Eq.\ref{evolved}. The
structure of each conformal set $\sigma_{I}, \sigma_{II},
\sigma_{III},...$ and consequently the iCF are preserved also if
we fix a different renormalization scale for each conformal
subset, i.e.:
\begin{eqnarray}
\left(\mu^2 \frac{ \partial}{\partial \mu^2} +\beta
(\alpha_s)\frac{\partial}{\partial \alpha_s}\right) \sigma_n=0.
\label{sigmainvariance}
\end{eqnarray}

We define here this property of Eq. \ref{confsubsets} of
separating an observable in the union of ordered scale invariant
disjoint subsets $\sigma_{I}, \sigma_{II}, \sigma_{III},...$ as
\textit{ordered scale invariance}.

In order to extend the iCF to all orders we first define a partial
limit $J_{/n}\rightarrow \infty$ as the limit obtained including
the higher order corrections relative only to those
$\beta_0,\beta_1,\beta_2,...,\beta_{n-2}$ terms that have been
determined already at the order $n$ for each subset and then we
perform the complementary $\bar{n}$ limit which consists in
including all the remaining terms. For the $J_{/n}$ limit we have:

\begin{eqnarray}
 \lim_{J_{/n}\rightarrow\infty}&
\sigma_I & \rightarrow
\left(\frac{\left. \alpha_{s}(\mu_{I})\right|_{n-2}}{2 \pi}\right) A_{\mathit{Conf}} \nonumber \\
\lim_{J_{/n}\rightarrow\infty}& \sigma_{II} & \rightarrow
\left(\frac{\left.\alpha_{s}(\mu_{II})\right|_{n-3}}{2
\pi}\right)^2  B_{\mathit{Conf}}  \nonumber \\
 \lim_{J_{/n}\rightarrow\infty}& \sigma_{III} & \rightarrow
\left(\frac{\left.\alpha_{s}(\mu_{\mathit{III}})\right|_{n-4}}{2
\pi}\right)^3
C_{\mathit{Conf}} \nonumber \\
 & \vdots & \qquad
\qquad \vdots \nonumber \\
\lim_{J_{/n}\rightarrow\infty}& \sigma_{n} & \equiv
 \left(\frac{\alpha_{s}(\mu_{0})}{2
\pi}\right)^n \mathcal{L}_{n \mathit{Conf}} \label{jnlimit}
\end{eqnarray}

where the $\left. \alpha_{s}(\mu_{I})\right|_{n-2}$ is the
coupling calculated up to the $\beta_{n-2}$ at the intrinsic scale
$\mu_I$. Given the particular ordering of the powers of the
coupling, in each conformal subset we have the coefficients of the
$\beta_0,...,\beta_{n-k-1}$ terms, where $k$ is the order of the
conformal subset and the $n$ is the order of the highest subset
with no $\beta$-terms. We notice that the limit of each conformal
subset is finite and scale invariant up to the $\sigma_{n-1}$. The
remaining scale dependence is confined in the coupling of the
$n^{th}$ term. Any combination of the $\sigma_I,...,\sigma_{n-1}$
subsets is finite and scale invariant. We can now extend the iCF
to all orders performing the $\bar{n}$ limit. In this limit we
include all the remaining higher order corrections. For the
calculated conformal subsets this leads to define the coupling at
the same scales but including all the missing $\beta$ terms. Thus
each conformal subset remains scale invariant. We point out that
we are not making any assumption on the convergence of the series
for this limit. Then we have:

\begin{eqnarray}
\lim_{\bar{n}\rightarrow\infty} & \sigma_I & \rightarrow
\left(\frac{ \alpha_{s}(\mu_{I})}{2 \pi}\right) A_{\mathit{Conf}} \nonumber \\
\lim_{\bar{n}\rightarrow\infty}  & \sigma_{II} &\rightarrow
\left(\frac{\alpha_{s}(\mu_{II})}{2
\pi}\right)^2  B_{\mathit{Conf}}  \nonumber \\
\lim_{\bar{n}\rightarrow\infty} & \sigma_{III} & \rightarrow
\left(\frac{\alpha_{s}(\mu_{\mathit{III}})}{2 \pi}\right)^3
C_{\mathit{Conf}} \nonumber \\
 & \vdots & \qquad
\qquad \vdots \nonumber \\
\lim_{\bar{n}\rightarrow\infty} & \sigma_{n} \!\! & \! \! \equiv
\!\! \lim_{n \rightarrow\infty} \left(\frac{\alpha_{s}(\mu_{0})}{2
\pi}\right)^n \!\! \mathcal{L}_{n \mathit{Conf}} \!\! \rightarrow \hbox{Conformal Limit} \nonumber \\
 & & \label{nlimit}
\end{eqnarray}
where here now $ \alpha_{s}(\mu_{I})$ is the complete coupling
determined at the same scale $\mu_I$. The Eq.\ref{nlimit} shows
that the whole renormalization scale dependence has been
completely removed. In fact both the intrinsic scales $\mu_N$ and
the conformal coefficients
$A_{\mathit{Conf}},B_{\mathit{Conf}},C_{\mathit{Conf}},...,\mathcal{L}_{n
\mathit{Conf}},...$ are not depending on the particular choice of
the initial scale. The only term with a residual $\mu_0$
dependence is the n-term, but this dependence cancels in the limit
$n\rightarrow \infty$. The scale dependence is totally confined in
the coupling $\alpha_s (\mu_0)$ and its behavior doesn't depend on
the particular choice of any scale $\mu_0$ in the perturbative
region, i.e. $\lim_{n\rightarrow\infty} \alpha_s(\mu_0)^n \sim
a^n$ with $a<1$.  Hence the limit of $\lim_{n\rightarrow\infty}
\sigma_{n}$ depends only on the properties of the theory and not
on the scale of the coupling in the perturbative regime.
 The proof given here shows that the iCF is {\it sufficient} to have a scale
invariant observable and it is not depending on the particular
convergence of the series. In order to show the {\it necessary}
condition we separate the two cases of a convergent series and an
asymptotic expansion. For the first case the {\it necessary}
condition stems directly from the uniqueness of the iCF form,
since given a finite limit and the scale invariance any other
parametrization can be reduced to the iCF by means of appropriate
transformations in agreement with the RG equations. For the second
case, we have that an asymptotic expansion though not convergent,
can be truncated at a certain order $n$, which is the case of
Eq.\ref{confsubsets}. Given the particular structure of the iCF we
can perform the first partial limit $J_{/n}$ and we would achieve
a finite and scale invariant prediction ,
$\sigma_{N-1}=\Sigma_{i=1}^{n-1} \sigma_i$, for a truncated
asymptotic expansion, as shown in Eq. \ref{jnlimit}. Given the
truncation of the series in the region of maximum of convergence
the n-th term would be reduced to lowest value and so the scale
dependence of the observable would reach its minimum. Given the
finite and scale invariant limit $\sigma_{N-1}$ we conclude that
the iCF is unique and then {\it necessary} for an {\it ordered}
scale invariant truncated asymptotic expansion up to the $n^{th}$
order. We point out that in general the iCF form is the most
general and irreducible parametrization which leads to the scale
invariance, other parametrization are forbidden since if we
introduce more scales\footnote{Here we refer to the form of
Eq.\ref{newevolution}. In principle it is possible to write other
parametrizations preserving the scale invariance, but these can be
reduced to the iCF by means of appropriate transformations in
agreement with the RG equations.} in the logarithms of one subset
we would spoil the invariance under the RG transformation and we
could not achieve Eq. \ref{evolved}, while on the other hand no
scale dependence can be introduced in the intrinsic scales since
it would remain in the observable already in the first partial
limit $J_{/n}$ and it could not be eliminated. The conformal
coefficients are conformal by definition at each order, thus they
don't depend on the renormalization scale and they don't have a
perturbative expansion. Hence {\it the iCF is a necessary and
sufficient condition for scale invariance.}

\subsection{The iCF and the ordered scale invariance }

The iCF-parametrization can stem either from an inner property of
the theory, the iCF, or from direct parametrization of the
scale-invariant observable. In both cases the iCF-parametrization
makes the scale dependence of the observable explicit and it
exactly preserves the scale invariance. Once we have defined an
observable in the iCF-form, we have not only the scale invariance
of the entire observable, but also the \textit{ordered scale
invariance} (i.e. the scale invariance of each subset $\sigma_n$
or $\sigma_{N-1}$). The latter property is crucial in order to
obtain scale invariant observables independently from the
particular kinematical region and independently from the starting
order of the observable or the order of the truncation of the
series. Since in general, a theory is blind respect to the
particular observable/process that we are going to investigate,
the theory should preserve the \textit{ordered} scale invariance
in order to define always scale invariant observables. Hence if
the iCF is an inner property of the theory, it leads to implicit
coefficients that are neither independent nor conformal. This is
made explicit in the Eq.\ref{newevolution}, but it is hidden in
the perturbative calculations in case of the implicit
coefficients. For instance the presence of the iCF clearly reveals
when a particular kinematical region is approached and the $A_O$
becomes null. This would cause a break of the scale invariance
since a residual initial scale dependence would remain in the
observable in the higher orders coefficients. The presence of the
iCF solves this issue by leading to the correct redefinition of
all the coefficients at each order preserving the correct scale
invariance exactly. Thus in the case of a scale invariant
observable $O$ defined, according to the implicit form
(Eq.\ref{observable1}), by the coefficients $ \{A_O, B_O,
C_O,..,O_O,...\}$ , it cannot simply undergo the change $
\rightarrow \{0 , B_O, C_O,..,O_O,...\}$ , since this would break
the scale invariance. In order to preserve the scale invariance we
must redefine the coefficients $\{\tilde{A}_O=0, \tilde{B}_O,
\tilde{C}_O,..,\tilde{O}_O,...\}$ cancelling out all the initial
scale dependence originated from the LO coefficient $A_O$ at all
orders. This is equivalent to subtract out a whole invariant
conformal subset $\sigma_I$ related to the coefficient
$A_{\mathit{Conf}}$ from the scale invariant observable $O$. This
mechanism is clear in the case of the explicit form of the iCF,
Eq. \ref{newevolution}, where if $A_{\mathit{Conf}}=0$ then the
whole conformal subset is null and the scale invariance is
preserved. We underline that the conformal coefficients can
acquire all the possible values without breaking the scale
invariance, they contain the essential information on the physics
of the process, while all the correlation factors can be
reabsorbed in the renormalization scales as shown by the PMC
method~\cite{PMC1,PMC2,PMC3,mojaza1,mojaza2}. Hence if a theory
has the property of the \textit{ordered scale invariance} it
preserves exactly the scale invariance of observables
independently from the process, the kinematics and the starting
order of the observable. We put forward that if a theory has the
Intrinsic Conformality all the renormalized quantities, such as
cross sections, can be parametrized  with the iCF-form. This
property should be preserved by the renormalization scheme or by
the definition of IR safe quantities and it should be preserved
also in observables defined in effective theories.


\subsection{The PMC$_\infty$}

We introduce here a new method to eliminate the scale-setting
ambiguity in single variable scale invariant distributions named
PMC$_\infty$. This method is based on the original PMC principle
~\cite{PMC1,PMC2,PMC3,mojaza1,mojaza2} and agrees with all the PMC
different formulations for the PMC-scales at the lowest order.
Essentially the core of the PMC$_\infty$ is the same of all the
BLM-PMC prescriptions\cite{blm}, i.e. the correct running coupling
value and hence its renormalization scale at the lowest order is
identified by the $\beta_0$-term at each order, or equivalently by
the \textit{intrinsic conformal scale} $ \mu_N$. The PMC$_\infty$
preserves the iCF and then the scale invariance absorbing an
infinite set of $\beta$-terms at all orders. This method differs
from the other PMC prescriptions since, due to the presence of the
Intrinsic Conformality, no perturbative correction in $\alpha_s$
needs to be introduced at higher orders in the PMC-scales. Given
that all the $\beta$-terms of a single conformal subset are
included in the renormalization scale already with the definition
at lowest order, no initial scale or scheme dependence are left
due to the unknown $\beta$-terms in each subset. The
PMC$_\infty$-scale of each subset can be unambiguously determined
by $\beta_0$-term of each order, we underline that all logarithms
of each subset have the same argument and all the differences
arising at higher orders have to be included only in the conformal
coefficients. Reabsorbing all the $\beta$-terms into the scale
also the $\beta_0^n n!$ terms (related to the renormalons
\cite{renormalons1}) are eliminated, thus the precision is
improved and the perturbative QCD predictions can be extended to a
wider range of values. The initial scale dependence is totally
confined in the unknown PMC$_\infty$ scale of the last order of
accuracy (i.e. up to NNLO case in the $\alpha_s(\mu_0)^3$).
 Thus if we fix the renormalization scale independently to
the proper intrinsic scale for each subset $\mu_N$, we end up with
a perturbative sum of totally conformal contributions up to the
order of accuracy:
\begin{eqnarray}
 \! \!\! \!  \! \!\! \!\frac{1}{\sigma_{0}} \frac{O d \sigma(\mu_I,\mu_{II},\mu_{III})}{d
O}\!\!\! &=&  \left\{
\frac{\alpha_{s}(\mu_I)}{2 \pi}   \frac{O dA_{\mathit{Conf}}}{d O}+ \! \! \right. \nonumber \\
 \! \!\! \!
\left(\frac{\alpha_{s}(\mu_{II})}{2 \pi}\right)^{2} \! \! \frac{O
d
 B_{\mathit{Conf}}}{d O} \! \! & + & \! \!
\left.  \left(\frac{\alpha_{s}(\mu_{III})}{2 \pi}\right)^{3} \! \!
\frac{O d
 C_{\mathit{Conf}}}{d O} \right\} \nonumber \\ & & \qquad  \qquad \qquad \,+ {\cal O}(\alpha_{s}^4),
 \label{observable2}
\end{eqnarray}
at this order $\mu_{III}=\mu_0$.

\subsection{iCF coefficients and scales}

We describe here how all the coefficients of Eq.
\ref{newevolution} can be identified from a numerical either
theoretical perturbative calculation. We will use as template the
NNLO thrust distribution results calculated in Refs.
\cite{weinzierl1,weinzierl2}. Since the leading order is already
($A_{\mathit{Conf}}$) void of $\beta$-terms we start with NLO
coefficients. A general numerical/theoretical calculation keeps
tracks of all the color factors and the respective coefficients:
\begin{eqnarray}
B_{O}(N_f)=C_F \left[ C_A B_{O}^{N_c}+C_F B_{O}^{C_F}+ T_F N_f
B_{O}^{N_f}\right] \label{Bcoeff}
\end{eqnarray}
where $C_F=\frac{\left(N_{c}^{2}-1\right)}{2 N_{c}}$, $C_A=N_c$
and $T_F=1/2.$ The dependence on $N_f$ is made explicit here for
sake of clarity. We can determine the conformal coefficient
$B_{\mathit{Conf}}$ of the NLO order straightforwardly, by fixing
the number of flavors $N_f$ in order to kill the $\beta_0$ term:
\begin{eqnarray}
B_{\mathit{Conf}}&=& B_{O} \left( N_f \equiv \frac{33}{2} \right),\nonumber \\
B_{\beta_0} \equiv  \log  \frac{\mu_0^2}{\mu_I^2}  & = & 2
\frac{B_O-B_{\mathit{Conf}}}{\beta_0 A_{\mathit{Conf}}}
\label{Bconf}
\end{eqnarray}
 we would achieve the same results in the usual PMC
way, i.e. identifying the $N_f$ coefficient with the $\beta_0$
term and then determining the conformal coefficient. Both methods
are consistent and results for the intrinsic scales and the
coefficients are in perfect agreement. At the NNLO a general
coefficient is made of the contribution of six different color
factors:
\begin{eqnarray}
C_{O}(N_f)&=& \frac{C_F}{4} \left\{ N_{c}^{2} C_{O}^{N_c^2
}+C_{O}^{N_c^0}+\frac{1}{N_{c}^{2}} C_{O}^{\frac{1}{N_c^2}}
 \right. \nonumber \\ & & \!\!\!\!\!\!\!\!\!\!\!\!\!\!\!\!\!\!\!\!\left.\!\!\!\!\!  +N_{f} N_{c}\cdot C_{O}^{N_f N_c}+ \frac{N_{f}}{N_{c}}
C_{O}^{N_f/N_c}+N_{f}^{2} C_{O}^{N_f^2}\right\}. \label{Ccoeff}
\end{eqnarray}
In order to identify all the terms of Eq.\ref{newevolution} we
notice first that the coefficients of the terms $\beta_0^2$ and
$\beta_1$ are already given by the NLO coefficient $B_{\beta_0}$,
then we need to determine only the $\beta_0$- and the conformal
$C_{\mathit{Conf}}$-terms. In order to determine the latter
coefficients we use the same procedure we used for the NLO , i.e.
we set the number of flavors $N_f \equiv 33/2$ in order to drop
off all the $\beta_0$ terms. We have then:
\begin{eqnarray}
C_{\mathit{Conf}}&=& C_{O} \left( N_f \equiv \frac{33}{2} \right)- \frac{1}{4}\overline{\beta}_1 B_{\beta_0} A_{\mathit{Conf}},\nonumber \\
C_{\beta_0} \equiv \log\left(\frac{\mu_0^2}{\mu_{II}^2}\right) & =
& \frac{1}{\beta_0 B_{\mathit{Conf}}} {\bigg (}
C_O-C_{\mathit{Conf}}  \nonumber \\
&\!\!\!\!  &\!\!\!\! -\left.\frac{1}{4} \beta_0^2 B_{\beta_0}^2
A_{\mathit{Conf}}- \frac{1}{4} \beta_1 B_{\beta_0}
A_{\mathit{Conf}} \!\right)\!\!,\nonumber \\
 \label{Cconf}
\end{eqnarray}
with $\overline{\beta}_1\equiv \beta_1(N_f=33/2)=-107$. This
procedure can be extended at every order and one may decide
whether to cancel the $\beta_0$, $\beta_1$ or $\beta_2$ by fixing
the appropriate number of flavors. The results can be compared
leading to determine exactly all the coefficients. We point out
that extending the Intrinsic Conformality to all orders we can
predict at this stage the coefficients of all the color factors of
the higher orders related to the $\beta$-terms except those
related to the higher order conformal coefficients and
$\beta_0$-terms (e.g. at NNNLO the $D_{\mathit{Conf}}$ and
$D_{\beta_0}$). The $\beta$-terms are coefficients that stem from
UV-divergent diagrams connected with the running of the coupling
constant and not from UV-finite diagrams. UV-finite $N_F$ terms
may arise but would not contribute to the $\beta$-terms. While
$N_f$ terms coming from UV-divergent diagrams, depending
dynamically on the virtuality of the underlying quark and gluon
subprocesses have to be considered as $\beta$-terms and they would
determine the intrinsic conformal scales. In general, each $\mu_N$
is an independent function of the $\sqrt{s}$, of the selected
variable $O$ and it varies with the number of colors $N_c$ mainly
due to $ggg$- and $gggg$-vertices. The latter terms arise at
higher orders only in non-Abelian theory but they do not spoil the
iCF-form. We underline that iCF applies to scale invariant single
variable distributions, in case one is interested in the
renormalization of a particular diagram, e.g. the $ggg$-vertex,
contributions from different $\beta$-terms should be singled out
in order to identify the respective intrinsic conformal scale
consistently with the renormalization of the non-Abelian
$ggg$-vertex, as shown in \cite{binger}.

\subsection{The PMC$_\infty$ scales at LO and NLO }

\begin{figure}[htb]
\centering
\includegraphics[width=0.40\textwidth]{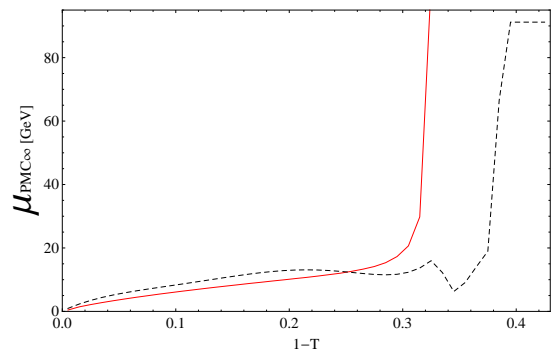}
\caption{The LO-PMC$_\infty$ (Solid Red) and the NLO-PMC$_\infty$
(Dashed Black) scales for thrust. } \label{Tscales}
\end{figure}

According to the PMC$_\infty$ prescription we fix the
renormalization scale to $\mu_N$ at each order absorbing all the
$\beta$ terms into the coupling. We notice a small mismatch
between the zeroes of the conformal coefficient
$B_{\mathit{Conf}}$ and those of the remainder $\beta_0$-term at
the numerator (formula is shown in Eq. \ref{Cconf}). Due to our
limited knowledge of the strong coupling at low energies, in order
to avoid singularities in the NLO-scale $\mu_{II}$, we introduce a
regularization which leads to a finite scale $\tilde{\mu}_{II}$.
These singularities stem from a rather logarithmic behavior of the
conformal coefficients when low values of the variable $1-T$ are
approached. Large logarithms arise from the IR divergences
cancellation procedure and they can be resummed in order to
restore a predictive perturbative
regime\cite{resummation0,resummation1,resummation2,resummation3,resummation5,resummation6}.
We point out that IR cancellation should not spoil the iCF
property and a IR cancellation Monte Carlo technique consistent
with the iCF would be required. Whether this is an actual
deviation from the iCF-form has to be further investigated.
 However since the discrepancies between the coefficients are rather small,
we introduce a regularization method based on redefinition of the
norm of the coefficient $B_{\mathit{Conf}}$ in order to cancel out
these singularities in the $\mu_{II}$-scale. This regularization
is consistent with the PMC principle and up to the accuracy of the
calculation it does not introduce any bias effect in the results
and no ambiguity in the NLO-PMC$_\infty$ scale. All the
differences introduced by the regularization would start at the
NNNLO order of accuracy and they can be absorbed after in the
higher order PMC$_\infty$ scales. Thus the first two PMC$_\infty$
scales result:

\begin{eqnarray}
\large{\mu}_{\mathit{I}} & = & \sqrt{s} \cdot e^{-\frac{1}{2} B_{\beta_0}},\hspace{1.9cm}{ \scriptstyle (1-T)<0.33}  \label{icfscale2} \\
\large{\tilde{\mu}}_{\mathit{II}} & =& \left\{ \begin{array}{lr}
\sqrt{s} \cdot e^{-\frac{1}{2} C_{\beta_0} \cdot
\frac{B_{\mathit{Conf}}}{B_{\mathit{Conf}}+\eta \cdot
A_{\mathit{tot}} A_{\mathit{Conf}} }},\\ \hspace{3.5cm} {\scriptstyle (1-T)<0.33}  \\
  \sqrt{s}\cdot e^{-\frac{1}{2}\left( \frac{C_1}{11
B_1-\frac{2}{3} B_0}\right)},\\ \hspace{3.5cm}{ \scriptstyle (1-T)>0.33} \label{icfscale}\\
 \end{array} \right.
 \label{PMC12}
\end{eqnarray}

where $\sqrt{s}=M_{Z_0}$, and the value of $\eta=3.51$ has been
fixed by matching the zeroes of numerator and denominator of
$C_{\beta_0}$. We have to point out that in the region
$(1-T)>0.33$ we have a clear example of Intrinsic Conformality-iCF
where the kinematical constraints set the $A_{\mathit{Conf}}=0$.
According to the Eq.\ref{evolved} setting the
$A_{\mathit{Conf}}=0$ the whole conformal subset $\sigma_I$
becomes null. In this case all the $\beta$ terms at NLO and NNLO
disappear except the  $\beta_0$-term at NNLO which determines the
$\mu_{II}$ scale. The surviving $N_f$ terms at NLO or the $N_f^2$
at NNLO are related to the finite $N_F$-term at NLO and to the
mixed $N_f \cdot N_F$ term arising from $B_O \cdot \beta_0$ at
NNLO. Using the following parametrization:
\begin{eqnarray}
 A_{O} &=&  0, \nonumber \\
B_{O}  &=& B_0+ B_1 \cdot N_f, \nonumber \\
 C_{O} &=& C_0+C_1 \cdot N_f +C_2 \cdot N_f^2.
  \label{Cparpar}
  \end{eqnarray}
 we can determine the $\tilde{\mu}_{II}$ for the region $(1-T)>0.33$ as shown in Eq.\ref{icfscale}:
\begin{equation}
 \tilde{\mu}_{II}=\sqrt{s} e^{-\frac{1}{2}\left( \frac{C_1}{11 B_1-\frac{2}{3}
 B_0}\right)}
  \label{mu2}
  \end{equation}
by identifying the $\beta_0$-term at NNLO. The LO and NLO
PMC$_\infty$ scales are shown in Fig.\ref{Tscales}.
We notice that the two PMC$_\infty$ scale have similar behaviors
in the range $(1-T)<0.33$ and the LO-PMC$_\infty$ scale agrees
with the PMC scale used in \cite{shengquan}. This method totally
eliminates both the ambiguity in the choice of the renormalization
scale and the scheme dependence at all orders in QCD.

\subsection{NNLO Thrust distribution results}

We use here the results of Ref. \cite{weinzierl1,weinzierl2} and
for the running coupling $\alpha_s(Q)$ we use the RunDec
program~\cite{rundec}. In order to normalize consistently the
thrust distribution we expand the denominator in $\alpha_0\equiv
\alpha_s(\mu_0)$ while the numerator has the couplings
renormalized at different PMC$_\infty$-scales $\alpha_I \equiv
\alpha_s(\mu_{I})$, $\alpha_{II}\equiv
\alpha_s(\tilde{\mu}_{II})$. We point out here that the proper
normalization would be given by the integration of the total cross
section after renormalization with the PMC$_\infty$ scales,
nonetheless since the PMC$_\infty$ prescription involves only
absorption of higher order terms into the scales the difference
would be within the accuracy of the calculations, i.e. $\sim
\mathcal{O}(\alpha_s^4(\mu_0))$. The experimentally measured
thrust distribution is normalized to the total hadronic cross
section $\sigma_{tot}$ as follows:

\begin{equation}
\frac{1}{\sigma_{tot}} \! \frac{O d
\sigma(\mu_I,\mu_{II},\mu_{0})}{d O}=
\left\{\overline{\sigma}_{I}+\overline{\sigma}_{II}+\overline{\sigma}_{III}+
{\cal O}(\alpha_{s}^4) \right\},
 \label{observable3}
\end{equation}
where $$ \sigma_{tot}=\sigma_{0}\left(1+\frac{\alpha_{s}(\mu_0)}{2
\pi} A_{t o t}+\left(\frac{\alpha_{s}(\mu_0)}{2 \pi}\right)^{2}
B_{t o t}+O\left(\alpha_{s}^{3}\right)\right)$$ is the total
integrated cross section and the $A_{\mathit{tot}},
B_{\mathit{tot}}$ are:
\begin{eqnarray}
A_{\mathit{tot}} &= & \frac{3}{2} C_F ; \\
B_{\mathit{tot}} &= & \frac{C_F}{4}N_c +\frac{3}{4}C_F
\frac{\beta_0}{2} (11-8\zeta(3)) -\frac{3}{8} C_F^2.
 \label{norm}
\end{eqnarray}
\newline The normalized subsets in
the region $(1-T)<0.33$ are then:
\begin{eqnarray}
\overline{\sigma}_{I} &=& A_{\mathit{Conf}} \frac{\alpha_I}{2 \pi} \nonumber  \\
\overline{\sigma}_{II} &=& \!\!\! \left( B_{\mathit{Conf}}+\eta
A_{\mathit{tot}} A_{\mathit{Conf}} \right)\!\!
\left(\frac{\alpha_{II}}{2 \pi} \right)^{\!2}
\!\!\! - \!\eta A_{\text{tot}} A_{\mathit{Conf}} \!\left(\frac{\alpha_0}{2 \pi}\!\right)^{\!2} \nonumber \\
 & & -A_{\text{tot}} A_{\mathit{Conf}} \frac{\alpha_0}{2 \pi} \frac{\alpha_I}{2 \pi}  \nonumber \\
\overline{\sigma}_{III} &=&\!\! \left( C_{\mathit{Conf}} \!-\!
A_{\text {tot}} \!B_{\mathit{Conf}}\!-\!(B_{\text
{tot}}\!-\!A_{\text {tot}}^{2}) A_{\mathit{Conf}}\! \right) \!\!
\left(\! \frac{\alpha_0}{2 \pi}\! \right)^{\!3}\nonumber \\
\label{normalizedcoeff}
\end{eqnarray}

Normalized subsets for the region $(1-T)>0.33$ can be achieved
simply by setting $A_{\mathit{Conf}}\equiv 0$ in the Eq.
\ref{normalizedcoeff}. Within the numerical precision of these
calculations there is no evidence of the presence of spurious
terms, such as UV-finite $N_f$ terms up to NNLO~\cite{monni}.
These terms must be rather small or comparable with numerical
fluctuations. Besides we notice a small rather constant difference
between the iCF-predicted and the calculated coefficient for the
$N_f^2$-color factor of Ref.~\cite{weinzierl1} which might be
addressed to a $N_f^2$ UV-finite coefficient or possibly to
statistics. This small difference must be included in the
conformal coefficient and it has a complete negligible impact on
the total thrust distribution.
\begin{figure}[htb]
\centering
\includegraphics[width=0.40\textwidth]{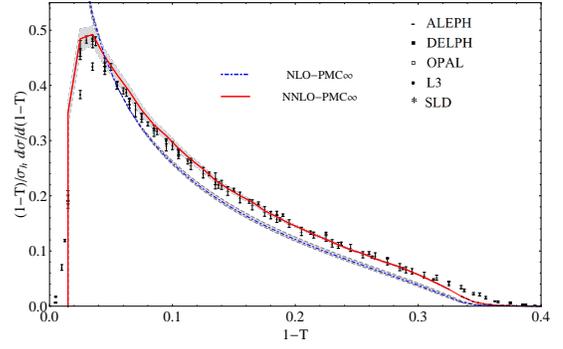}
\caption{The thrust distribution under the PMC$_\infty$ at NLO
(DotDashed Blue) and at NNLO (Solid Red). The experimental data
points are taken from the ALEPH, DELPHI,OPAL, L3, SLD experiments
\cite{aleph,delphi,opal,l3,sld}. The shaded area shows theoretical
errors for the PMC$_\infty$ predictions at NLO and at NNLO.}
\label{thrust}
\end{figure}
 In Fig.\ref{thrust} we show the thrust
distribution at NLO and at NNLO with the use of the PMC$_\infty$
method. Theoretical errors for the thrust distribution at NLO and
at NNLO are also shown (the shaded area). Conformal quantities are
not affected by a change of renormalization scale. Thus the errors
shown give an evaluation of the level of conformality achieved up
to the order of accuracy and they have been calculated using
standard criteria, i.e. varying the remaining initial scale value
in the range $\sqrt{s}/2 \leq \mu_0 \leq 2 \sqrt{s}$. Using the
same definition of the parameter $\bar{\delta}$ as in
Ref.~\cite{gehrmannthrust}, we have in the interval
$0.<(1-T)<0.33$ an average error of $\bar{\delta}\simeq 3.54\%$
and $1.77\%$ for the thrust at NLO and at NNLO respectively. A
larger improvement has been calculated in the range
$0.<(1-T)<0.42$ from $\bar{\delta}\simeq 7.36\%$ to $1.95\%$ from
NLO to the NNLO accuracy with the PMC$_\infty$.

\begin{figure}[htb]
\centering
\includegraphics[width=0.40\textwidth]{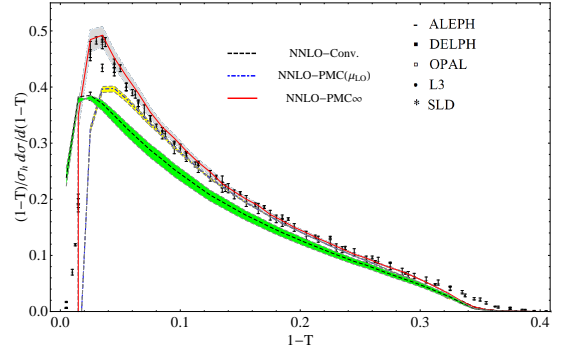}
\caption{The thrust distribution at NNLO under the Conventional
(Dashed Black), the PMC($\mu$\textsubscript{LO}) (DotDashed Blue)
and the PMC$_\infty$ (Solid Red). The experimental data points are
taken from the ALEPH, DELPHI,OPAL, L3, SLD experiments
\cite{aleph,delphi,opal,l3,sld}.  The shaded areas show
theoretical errors predictions at NNLO and they have been
calculated varying the remaining initial scale value in the range
$\sqrt{s}/2 \leq \mu_0 \leq 2 \sqrt{s}$.} \label{thrust2}
\end{figure}

In Fig.\ref{thrust2} a direct comparison of the PMC$_\infty$ with
the the Conventional Scale Setting results (obtained in
\cite{weinzierl1} and \cite{gehrmannthrust}\cite{gehrmannshapes1})
is shown. In addition we have shown also the results of the first
PMC approach used in \cite{shengquan} that we indicate as
PMC($\mu$\textsubscript{LO}) extended to the NNLO accuracy. In
this approach the last unknown PMC scale $\mu$\textsubscript{NLO}
of the NLO has been set to the last known PMC scale
$\mu$\textsubscript{LO} of the LO, while the NNLO scale
$\mu$\textsubscript{NNLO}$\equiv \mu_0$ has been left unset and
varied in the range $\sqrt{s}/2 \leq \mu_0 \leq 2 \sqrt{s}$. This
analysis has been performed in order to show that the procedure of
setting the last unknown scale to the last known one leads to
stable and precise results and is consistent with proper PMC
method in a wide range of values of the $(1-T)$ variable.

\begin{table}[h!]
\centering
 \begin{tabular}{||c|c|c|c||}
   \hline
 $\bar{\delta}[\%]$  &  Conv.  & PMC($\mu$\textsubscript{LO}) & PMC$_\infty$ \\
    \hline
   $0.10 < (1-T) < 0.33$ & 6.03 & 1.41 & 1.31 \\
    $0.21 < (1-T) < 0.33$ & 6.97 & 2.19 & 0.98 \\
    $0.33 < (1-T) < 0.42$ & 8.46 & -    & 2.61 \\
    $0.00 < (1-T) < 0.33$ &  5.34  & 1.33 & 1.77\\
   $0.00 < (1-T) <0.42$ & 6.00 & -  &  1.95 \\
   \hline
   \end{tabular}
 \caption{Average error, $\bar{\delta}$, for NNLO Thrust distribution under Conventional, PMC($\mu$\textsubscript{LO}) and PMC$_\infty$
 scale settings calculated in different ranges of values of the $(1-T)$ variable.}
 \label{tab:1}
\end{table}

Average errors calculated in different regions of the spectrum are
reported in Table \ref{tab:1}. From the comparison with the
Conventional Scale Setting we notice that the PMC$_\infty$
prescription significantly improves the theoretical predictions.
Besides, results are in remarkable agreement with the experimental
data in a wider range of values ( $0.015 \leq 1-T \leq 0.33$) and
they show an improvement of the PMC$(\mu_{LO})$ results when the
two-jets and the multi-jets regions are approached, i.e. the
region of the peak and the region $(1-T)>0.33$ respectively. The
use of the PMC$_\infty$ approach on perturbative thrust
QCD-calculations restores the correct behavior of the thrust
distribution in the region $(1-T)>0.33$ and this is a clear effect
of the iCF property. Comparison with the experimental data has
been improved all over the spectrum and the introduction of the
$N^3LO$ order correction would improve this comparison especially
in the multi-jet $1-T > 0.33$ region. In the PMC$_\infty$ method
theoretical errors are given by the unknown intrinsic conformal
scale of the last order of accuracy. We expect this scale not to
be significantly different from that of the previous orders. In
this particular case, as shown in Eq.\ref{normalizedcoeff}, we
have also a dependence on the initial scale $\alpha_s(\mu_0)$ left
due to the normalization and to the regularization terms. These
errors represent the 12.5\% and 1.5\% respectively of the whole
theoretical errors in the range $0<(1-T)<0.42$ and they could be
improved by means of a correct normalization.

\subsection{NNLO C-parameter distribution results}

\begin{figure}[ht]
\centering
\includegraphics[width=0.40\textwidth]{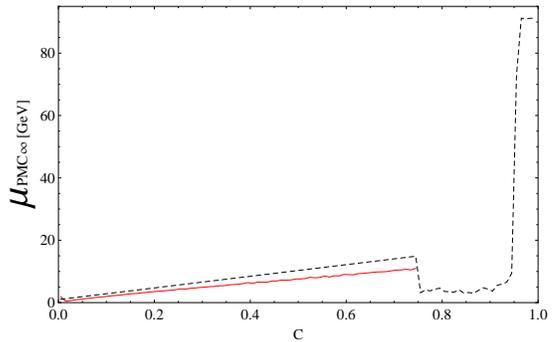}
\caption{The LO-PMC$_\infty$ (Solid Red) and the NLO-PMC$_\infty$
(Dashed Black) scales for C-parameter.} \label{Cpar-scales}
\end{figure}

\begin{figure}[ht]
\centering
\includegraphics[width=0.40\textwidth]{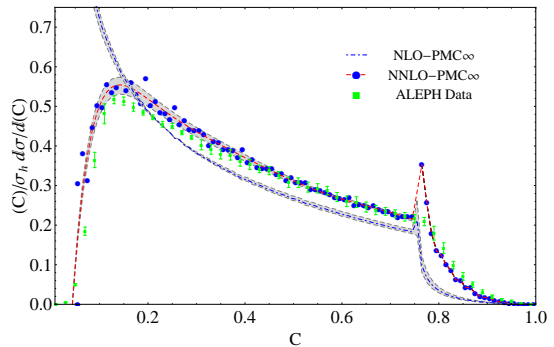}
\caption{The C-parameter distribution under the PMC$_\infty$ at
NLO (DotDashed Blue) and at NNLO (Dashed Red). Blue points
indicate the NNLO-PMC$_\infty$ thrust distribution obtained with
$\mu_{III}=\mu_0=M_{Z_0}$. The experimental data points (Green)
are taken from the ALEPH experiment~\cite{aleph}. Dashed lines of
the NNLO distribution show fits of the theoretical calculations
with interpolating functions for the values of the remaining
initial scale $\mu_0 = 2 M_{Z_0}$ and  $M_{Z_0}/2$. The shaded
area shows theoretical errors for the PMC$_\infty$ predictions at
NLO and at NNLO calculated varying the remaining initial scale
value in the range $\sqrt{s}/2 \leq \mu_0 \leq 2 \sqrt{s}$.}
\label{Cpar}
\end{figure}

The same analysis applies straightforwardly to the C-parameter
distribution including the regularizing $\eta$ parameter which has
been set to the same value $3.51$. The same scales of
Eq.\ref{icfscale2} and Eq.\ref{icfscale} apply to the C-parameter
distribution in the region $0<C<0.75$ and in the region
$0.75<C<1$. In fact, due to kinematical constraints that set the
$A_{\mathit{Conf}}=0$, we have the same iCF effect also for the
C-parameter. Results for the C-parameter scales and distributions
are shown in Fig.\ref{Cpar-scales} and Fig.\ref{Cpar}
respectively.

Theoretical errors have been calculated, as in the previous case,
using standard criteria and results indicate an average error over
the whole spectrum $0<(C)<1$ of the C-parameter distribution at
NLO and at NNLO of $\bar{\delta}\simeq 7.26\%$ and $2.43\%$
respectively.

\begin{table}[htb]
\centering
 \begin{tabular}{ || c |c|c|c||  }
   \hline
 $\bar{\delta}$[\%]  &  Conv.  & PMC($\mu$\textsubscript{LO}) & PMC$_\infty$ \\
   \hline
   $0.00 < (C) < 0.75$ & 4.77 & 0.85 & 2.43 \\
    $0.75 < (C) < 1.00$ & 11.51 & 3.68 & 2.42 \\
    $0.00 < (C) < 1.00$ & 6.47 & 1.55 & 2.43 \\
  \hline
 \end{tabular}
 \caption{Average error  ,$\bar{\delta}$, for NNLO C-parameter distribution under Conventional, PMC($\mu$\textsubscript{LO}) and PMC$_\infty$
 scale settings calculated in different ranges of values of the $(C)$ variable.}
 \label{tab:2}
\end{table}

A comparison of average errors according to the different methods
is shown in Table\ref{tab:2}. Results show that the PMC$_\infty$
improves the NNLO QCD predictions for the C-parameter distribution
all over the spectrum.

\begin{figure}[h]
\centering
\includegraphics[width=0.40\textwidth]{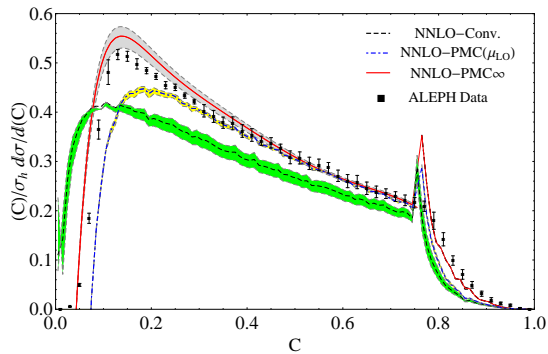}
\caption{The NNLO C-parameter distribution under the Conventional
Scale Setting (Dashed Black), the PMC($\mu$\textsubscript{LO})
(DotDashed Blue ) and the PMC$_\infty$ (Solid Red). The
experimental data points (Black) are taken from the ALEPH
experiment~\cite{aleph}. The shaded area shows theoretical errors
predictions at NNLO calculated varying the remaining initial scale
value in the range $\sqrt{s}/2 \leq \mu_0 \leq 2 \sqrt{s}$.}
\label{Cpar2}
\end{figure}

Comparison of the distributions calculated with the Conventional
Scale Setting, the PMC($\mu$\textsubscript{LO}) and the
PMC$_\infty$ is shown in Fig.\ref{Cpar2}. Results for the
PMC$_\infty$ show a remarkable agreement with the experimental
data away from the regions $C<0.05$ and $C\simeq 0.75$. The errors
due to the normalization and to the regularization terms
(Eq.\ref{normalizedcoeff}) are respectively the $8.8\%$ and
$0.7\%$ of the whole theoretical errors. The perturbative
calculations could be further improved using a correct
normalization and also by introducing the large logarithms
resummation technique in order to extend the perturbative regime.

\subsection{Comment on the last PMC$_\infty$ scale}

We have shown in this article a property of the perturbative QCD
corrections which is consistent with the MC results of single
variable distributions for $e^+e^-\rightarrow 3 jets$. This
property leads to the PMC$_\infty$ : an infinite-order scale
setting based on the Principle of Maximum Conformality method. The
PMC$_\infty$ method preserves the iCF form and is void of
ambiguities. The absence of ambiguities in PMC$_\infty$ is clearly
shown hereby simply noticing that the new scale at a each order is
fixed by considering the $\beta_0$- i.e. $N_f$-term and no other
term is needed in agreement with the iCF parametrization. Thus
PMC$_\infty$ scale fixing procedure is constrained by the iCF form
and the choice of the scales is totally free of any ambiguity. The
only unknown scale which remains unfixed apparently, is the one
given by the last order of accuracy. In this article this scale
has been fixed to the $\mu_{III}=\mu_0=M_{Z_0}$ for sake of a
consistent comparison with the Conventional method.
We remark that the last "unknown" PMC scale can be fixed to the
last known one. In fact, as we have also shown in this article, in
general the differences between two consecutive scales are not
only rather small, but the use of the method
PMC($\mu$\textsubscript{LO}) leads to precise and consistent
results in a wide range of the perturbative region as reported in
Table \ref{tab:1} and \ref{tab:2} of the previous sections.
 In the case of the thrust and C-parameter distributions calculated at
NNLO with the PMC$_\infty$, the differences between the
$\mu_{III}=M_{Z_0}$ or $\mu_{III}=\tilde{\mu}_{II}$ are also
negligible and these differences are totally confined within the
errors shown for the distributions in the perturbative region.
Given the {\it Conformal Limit} of the Eq.~\ref{nlimit}, the last
term in the iCF determines the level of {\it conformality} reached
by the expansion and the whole scale dependence is confined in the
coupling. As previously shown, this term is the main source of the
uncertainties (as also reported in the
Ref.\cite{Chawdhry:2019uuv}) and a further investigation regarding
the regularization method applied here would be necessary
in order to extend the procedure to all orders.

\section{Conclusions}

We have introduced here the iCF \textit{Intrinsic Conformality}, a
parametrization which preserves exactly the scale invariance of an
observable. We have shown a new method to solve the conventional
renormalization scale ambiguity in QCD named PMC$_\infty$ which
preserves the iCF and leads to conformal renormalization scales.
This method agrees with the PMC at LO and and it applies to
abservables especially in cases of a clearly manifested iCF, as
for example in the case of the event shape variables. We have
presented a new procedure to identify the
iCF/PMC$_\infty$-coefficients at all orders which can be applied
to either numerical or analytical calculations. The PMC$_\infty$
has been applied to the NNLO thrust and C-parameter distributions
and the results show perfect agreement with the experimental data.
The evaluation of theoretical errors using standard criteria show
that the PMC$_\infty$ significantly improves the theoretical
predictions all over the spectra of the shape variables.
 The PMC$_\infty$ method eliminates the renormalon growth $\alpha_s^n
\beta_0^n n!$, the scheme dependence and all the uncertainties
related to the scale ambiguity up to the order of accuracy. The
iCF/PMC$_\infty$ scales are identified by the lowest order
logarithm related to the $\beta_0$-term at each order and all the
physics of the process is contained in the conformal coefficients.
This is in complete agreement with QED and with the Gell-Mann and
Low scheme~\cite{qed,huet}.
 We have discussed why iCF should be considered a strict requirement for a
theory in order to preserve the scale invariance of the
observables and we have shown that iCF is consistent with the
single variable thrust and C-parameter distributions. We point out
that other conformal aspects of QCD resulting from different
sectors such as Commensurate Scale Relations-CSR~\cite{csr} or
dual theories as AdS/CFT~\cite{ads1} might also be related to the
Intrinsic Conformality.
 We underline that the iCF property in a theory would have two main remarkable consequences:
 first, it shows what is the correct coupling constant at each order as a function of the conformal intrinsic scale $\mu_N$,
and second, since only the conformal and the $\beta_0$
coefficients need to be identified in the observables at each
order, by means of the PMC$_\infty$ method the iCF would reveal
its predictive feature for the coefficients of the higher order
color factors. We point out that in many cases the implementation
of the iCF in a multi-loop calculations procedure would lead to a
significant reduction of the color factors coefficients and it
would speed up the calculations for higher order corrections.

\hspace{1cm}
 {\bf Acknowledgements}: We thank Francesco Sannino
for useful discussions. L.D.G. wants to thank the SLAC theory
group for its kind hospitality and support. This work was also
supported by the Department of Energy Contract No.
DE-AC02-76SF00515.\newline SLAC-PUB-17511.


\end{document}